\documentclass[aps,pre,showpacs,showkeys,twocolumn,superscriptaddress,
              floats,floatfix,preprintnumbers]{revtex4}
\usepackage{amssymb,amsmath}
\usepackage{bm,url,graphics,subfigure,epsfig,rotating,float}
\usepackage{ulem}
\usepackage{color,soul}
\def \AA {\mathcal{A}}
\def \Pphi {P_{\phi}}

\def \PI {P^{\rm I}}

\def \QI {Q^{\rm I}}

\def \TA {T_{\rm A}}
\def \TB {T_{\rm B}}
\def \TC {T_{\rm C}}
\def \TD {T_{\rm D}}
\def \TV {T_{\rm vortical}}
\def \TS {T_{\rm strain}}
\def \rhof {\rho_{\rm f}}

\def \uu  {{\bm u}}

\def \ff  {{\bm f}}

\def  \XX  {{\bm X}}
\def  \VV  {{\bm V}}

\def \taup {\tau_{\rm p}}

\def \grad {{\bm \nabla}}

\def \dive {{\bm \nabla}\cdot}
\def \lap {\nabla^2}
\def \delt {\partial_t}

\def \Teddy {T_{\rm eddy}}

\def \tper {\mathrm{t_{pers}}}

\def \Np  {N_{\rm p}}

\def \St  {{\rm St}}

\newcommand{\Fig}[1]{Fig.~(\ref{#1})}

\newcommand{\bfig}{\begin{figure}}
\newcommand{\efig}{\end{figure}}
\newcommand{\bc}{\begin{center}}
\newcommand{\ec}{\end{center}}
\newcommand{\bea}{\begin{eqnarray}}
\newcommand{\eea}{\end{eqnarray}}

\begin{document}
\title{How long do particles spend in vortical regions in turbulent flows?}
\author{Akshay Bhatnagar}
\email{akshayphy@gmail.com}
\affiliation{Centre for Condensed Matter Theory, Department of Physics, Indian Institute of Science, Bangalore 560012, India.}
\affiliation{Nordita, KTH Royal Institute of Technology and
Stockholm University, Roslagstullsbacken 23, 10691 Stockholm, Sweden}
\author{Anupam Gupta}
\email{anupam@physics.iisc.ernet.in}
\affiliation{Laboratoire de Génie Chimique, Universite de Toulouse, 
INPT-UPS, 31030, Toulouse, France.}
\author{Dhrubaditya Mitra}
\email{dhruba.mitra@gmail.com}
\affiliation{Nordita, KTH Royal Institute of Technology and
Stockholm University, Roslagstullsbacken 23, 10691 Stockholm, Sweden}
\author{Rahul Pandit}
\email{rahul@physics.iisc.ernet.in}
\affiliation{Centre for Condensed Matter Theory, Department of Physics, Indian Institute
of Science, Bangalore 560012, India.}
\author{Prasad Perlekar}
\email{perlekar@tifrh.res.in}
\affiliation{TIFR Centre for Interdisciplinary Sciences, 21 Brundavan Colony,
Narsingi, Hyderabad 500075, India}

\begin{abstract}
\end{abstract}
\preprint{NORDITA-2016-87}
\begin{abstract}
We obtain the probability distribution functions (PDFs) of the time that a
Lagrangian tracer or a heavy inertial particle spends in vortical or
strain-dominated regions of a turbulent flow, by carrying out direct numerical
simulation (DNS) of such particles advected by statistically steady,
homogeneous and isotropic turbulence in the forced, three-dimensional,
incompressible Navier-Stokes equation. We use the two invariants, $Q$ and $R$,
of the velocity-gradient tensor to distinguish between vortical and
strain-dominated regions of the flow and partition the $Q-R$ plane into four
different regions depending on the topology of the flow; out of these four
regions two correspond to vorticity-dominated regions of the flow and two
correspond to strain-dominated ones. We obtain $Q$ and $R$ along the
trajectories of tracers and heavy inertial particles and find out the time
$\tper$ for which they remain in one of the four regions of the $Q-R$ plane. We
find that the PDFs of $\tper$ display exponentially decaying tails for all four
regions for tracers and heavy inertial particles. From these PDFs we extract
characteristic times scales, which help us to quantify the time that such
particles spend in vortical or strain-dominated regions of the flow.   
\end{abstract}
\maketitle
\section{Introduction}

The characterization of the statistical properties of particles advected by a
turbulent flow is a challenging problem.  Not only is it of fundamental
interest  in fluid mechanics and non-equilibrium statistical mechanics, but it
also has applications in geophysical fluid dynamics (e.g., raindrop formation
in warm clouds~\cite{sha03,gra+wan13, pinsky1997turbulence,fal+fou+ste02}) and
astrophysics (e.g., planet formation in astrophysical disks
~\cite{Arm10,de2015planetary}). An important challenge here is to obtain the
time  that such advected particles spend in vortical regions of the flow.
We build on our studies of persistence-time statistics in two-dimensional (2D)
fluid turbulence~\cite{perlekar2011persistence} to develop a natural way of
defining a time for which a particle stays in a vortical region
in the three-dimensional (3D) case.
We illustrate
how this is done for the case of statistically steady, homogeneous, and
isotropic fluid turbulence by studying turbulent advection of (a) neutrally
buoyant Lagrangian tracers (henceforth called tracers), which move with the
fluid velocity at the particle, and (b) passive, heavy, inertial particles
(henceforth heavy particles), which are spherical particles that are heavier
than the carrier fluid and smaller than the Kolmogorov length scale $\eta$, at
which viscous dissipation becomes significant.  
The trapping of a tracer into a vortical region is expected to give rise
to very high values of particle acceleration~~\cite{toschi2005acceleration,biferale2005joint}.
The heavy particles are ejected from
vortices~\cite{eaton1994preferential,collins2004reynolds,yoshimoto2007self,bec2007heavy,TOSCHI,maxXX} hence they are preferentially 
found in strain-dominated regions of the flow. This has been observed 
in direct numerical simulations (DNSs) by overlaying the positions
of these particle on a pseudo-color plot of the magnitude of the vorticity, in a
two-dimensional slice~\cite{biferale2007inertial,bec2014gravity,
cencini2006dynamics} through the simulation domain. 

We estimate the time that a tracer or a heavy particle spends in a
vortical or strain-dominated region of the flow by using the following,
well-established technique for distinguishing between these flow
regions~\cite{chong1990general,cantwell1993behavior,perry1987description}:  At
any point in the flow, the velocity-gradient matrix $\AA$ has two invariants
$Q$ and $R$~\cite{chong1990general,cantwell1993behavior} (in the incompressible
case, that we consider, the trace of $\AA$ is zero everywhere).  Depending upon the
signs of $R$ and $\Delta = (27/4)R^2 + Q^3$, we can divide the $Q-R$ plane into
four regions (Fig.~\ref{fig:qr}); in two of these regions two eigenvalues of $\AA$ are complex
conjugates of each other; and the topology of the local flow is vortical.  The
other two regions of the $Q-R$ plane corresponds to those points for which all
the three eigenvalues of $\AA$ are real, and the local flow is
strain-dominated. In our DNSs, we follow the trajectories of tracers or heavy 
particles in time and calculate the velocity-gradient matrix $\AA$ at the
positions of these particles. The signs of $R$ and $\Delta$ help us to identify
whether a particle lies in a vortical or a strain-dominated
region of the flow at a given instant of time.  To obtain statistics for the
time scales over which such particles stay in vortical or strain-dominated
regions of the flow, it is natural to use the following idea of persistence
from non-equilibrium statistical mechanics:  For a
fluctuating field $\phi$, we find the probability distribution function (PDF)
$\Pphi(\tper)$, which gives the probability that $\phi$ does not change sign up
to time $\tper$. Persistence times can also be thought of as  
first-passage times~\cite{Redner}. 

Persistence has been studied in many non-equilibrium systems, e.g., the simple
diffusion equation with random initial conditions ~\cite{majumdar1996survival},
reaction-diffusion systems~\cite{ben1996reaction}, and fluctuating
interfaces~\cite{krug1997persistence}. In many systems it has been found that
$\Pphi(\tper) \sim \tper^{-\theta}$, as $\tper \to \infty$, where $\theta$ is
called the persistence exponent~\cite{bray2013persistence}.  This exponent
$\theta$ can be universal; it can be obtained analytically only in a few cases;
most often it is calculated numerically. We refer the reader to
Refs.~\cite{bray2013persistence,majumdar1999persistence} for reviews of such
persistence problems.   

In our DNS we calculate the PDF $\Pphi(\tper)$ of the times $\tper$ for which
tracers or heavy particles remain in vortical or strain-dominated region. We
find that, in the frame of tracers or heavy particles, these PDFs show
exponentially decaying tails, from which we extract the decay times scales. Our
study quantifies the dependence of these time scales on the Stokes number
$\St=\taup/\tau_\eta$, with $\taup$ the particle-response or Stokes time and
$\tau_\eta$ the dissipation-scale time and provides, therefore, a natural way
of answering the following question: How long do particles spend in vortical
regions in turbulent flows? 

The remainder of this paper is organized as follows. In Sec.~\ref{model} we
present the 3D Navier-Stokes equation, the equations we use for the time
evolution of tracers and heavy particles, and the numerical methods we use to
solve these; in subsection~\ref{qr} we define the two invariants $Q$ and $R$,
which we use to distinguish between vortical and strain-dominated regions of
the flow. Section~\ref{res} is devoted to a detailed description of our
results; and Section~\ref{conc} contains concluding remarks.

\section{Model and Numerical Methods}
\label{model}
We perform a DNS of the incompressible, three-dimensional,
forced, Navier-Stokes (3D NS) equation  
\begin{eqnarray}
\delt \uu + \uu\cdot \grad \uu &=& \nu \lap \uu - \grad p + \ff, \label{ns} \\
\dive \uu &=& 0, 
\label{incom}
\end{eqnarray}
where $\uu$, $p$, $\ff$, and $\nu$ are the velocity, pressure, force, and
kinematic viscosity, respectively.  Our simulation domain is a periodic box of
length $2\pi$.  We solve the 3D NS equation by using the
pseudo-spectral method with  $N^3$ collocation points and the $2/3$-dealiasing
rule~\cite{spectral}.  We use a constant-energy-injection forcing scheme
~\cite{Ganpati}, with a rate of energy injection $\varepsilon$. For time integration we
use a second-order, exponential  Adams--Bashforth scheme~\cite{akshaythesis}.

Heavy particles obey the following equations~\cite{gatignol1983faxen,maxey1983equation}:
\begin{eqnarray}
\dot{\XX} &=& \VV , \nonumber \\
\dot{\VV} &=& \frac{1}{\taup} \left[\uu(\XX) - \VV \right],
\label{eq:part}
\end{eqnarray}
where $\XX$ and $\VV$ denote, respectively, the position and velocity of the
particle, $\taup$ is the particle-response time, $\uu(\XX)$ is the flow
velocity at the position $\XX$, and dots denote time differentiation.  We
consider mono-disperse spherical particles, with radii $r_{\rm p} \ll \eta$,
material density $\rho_{\rm p}$ much greater than the fluid density $\rhof$,
and a small number density, so we neglect (a) the effect of the particles on
the flow (i.e., we have passive particles) and (b) particle-particle interactions.  We
also assume that, as in several experiments, typical particle accelerations, in
strongly turbulent flows, exceed significantly the acceleration because of
gravity. 
We also study the statistics of tracers for which
the equation of motion is 
\begin{equation}
\dot{\XX} = \uu(\XX).
\label{eq:tracer}
\end{equation}
We solve Eqs.~(\ref{eq:part}) and ~(\ref{eq:tracer}) by using an Euler scheme in time
to follow the trajectories
of $\Np$ particles in our DNS. 
The velocity-gradient matrix $\AA$ is calculated at each grid point
by using spectral method. 
We use trilinear interpolation to calculate the components of $\uu(\XX)$ and 
$\AA$ at the  off-grid positions of the particles.  
Table \ref{table:para} gives the list of
parameters we use in our DNS.

\begin{table}
\begin{center}
\caption{Table of parameters for our DNS run with $N^3$ collocation points:
$\nu$ is the kinematic viscosity, $\delta t$ the time step, $N_p$ is the number
of tracers or heavy particles, $k_{max}$ the largest wave number, $\epsilon$
the mean rate of energy dissipation; $\eta = (\nu^3/\epsilon)^{1/4}$ and
$\tau_\eta = (\nu/\epsilon)^{1/4}$ are the dissipation length and time scales,
respectively; $\lambda=\sqrt{2\nu E/\epsilon}$ is the Taylor micro-scale, where
$E$ is the mean energy of the flow, and  $Re_\lambda$ is the Reynolds number
based on $\lambda$, $I_l = \frac{\sum_k E(k)/k}{E}$ is integral length scale,
where $E(k)$ is the energy spectrum of the flow, and $T_{eddy} = I_l/u_{rms}$
is the large eddy turn-over time, where $u_{rms}$ is the root-mean-squared
velocity of the flow.}
\begin{tabular}{ c c c c c c }
\hline\hline
$N$ & $\nu$ & $\delta t$ & $N_p$ & $Re_\lambda$ & $k_{max}\eta$ \\
$256$ & $3.8\times 10^{-3}$ & $5\times10^{-4}$ & $40,000$ & $43$ & $1.56$ \\
\hline
$\epsilon$ & $\eta$ & $\lambda$ & $I_l$ & $\tau_\eta$ & $T_{eddy}$ \\
$0.49$ & $1.82\times10^{-2}$ & $0.16$ & $0.51$ & $8.76\times10^{-2}$ & $0.49$ \\
\hline
\end{tabular}
\end{center}
\label{table:para}
\end{table}

\subsection{$Q-R$ invariants of the velocity-gradient tensor}
\label{qr}
\begin{figure}
\begin{center}
\includegraphics[width=1.0\linewidth]{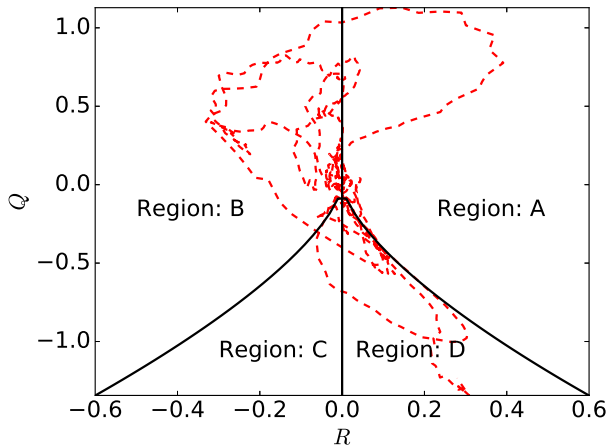}
\caption{(Color online) The flow is topologically different for values of $Q$
and $R$ that lie in the four regions shown in the $Q-R$ plane (after
Ref.~\cite{cantwell1993behavior}); the black curve is the zero-discriminant line
$\Delta = 0$. Regions A and B are vorticity-dominated regions; in region A
vortices are stretched and in region B they are compressed. By contrast,
regions C and D corresponds to strain-dominated or extensional regions; in
region C fluid elements experience biaxial strain, whereas, in region D, they
feel axial strain. The red dashed curve shows a illustrative path, in the
$Q-R$ plane, as a tracer moves through the fluid in our DNS.}
\label{fig:qr}
\end{center}
\end{figure}

We follow Ref.~\cite{cantwell1993behavior} to note that the velocity-gradient
matrix $\AA$ has three invariants under canonical transformations, namely,
$P=Tr(\AA)$, $Q=-Tr(\AA^2/2)$, and $R=-Tr(\AA^3/3)$.  Incompressibility yields
$P=0$, for all the points in our domain.  The nature of the eigenvalues is
determined by the signs of $R$ and $\Delta = (27/4)R^2 + Q^3$, the discriminant
of the characteristic equation of $\AA$.  This allows us to classify each point
in our flow into four regions, in the $Q-R$ plane, as shown in
Fig.~\ref{fig:qr}.  If $\Delta$ is large and positive, vorticity dominates the
flow; if, in addition, $R < 0$ (Region B), vortices are compressed, whereas, if
$R > 0$ (Region A), they are stretched. If $\Delta$ is large and negative,
local strains are high and vortex formation is not favored; furthermore, if $R
> 0$ (Region D), fluid elements experience axial strain, whereas, if $R < 0$
(Region C), they feel biaxial strain~\cite{cantwell1993behavior}.    

\begin{figure}
\begin{center}
\includegraphics[width=1.0\linewidth]{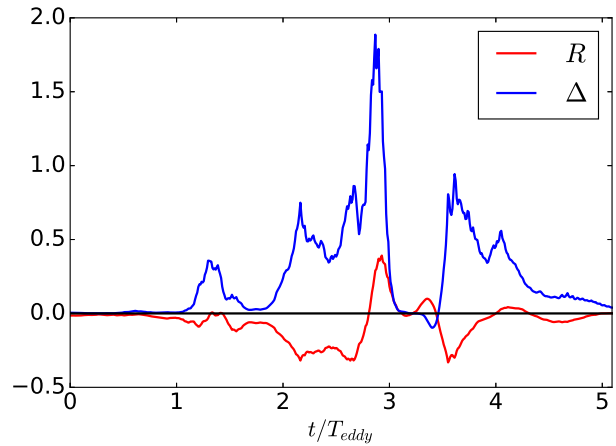}
\caption{(Color online) Plots of $R$ (blue) and the discriminant $\Delta$ of
the characteristic equation for the velocity-gradient tensor (green),
calculated along the trajectory of a tracer as a function of the
dimensionless time $t/T_{eddy}$. The intersection of any one of these curves
with the black horizontal line indicates the migration of a particle from one
region of the $Q-R$ plane to another. $R$ and $\Delta$ are 
Nondimensionalized by $\Lambda^3$ and $\Lambda^6$, respectively, where
$\Lambda=u_\eta/\eta$.}
\label{fig:timqr}
\end{center}
\end{figure}
\section{Results}
\label{res}

From our simulations we find that the iso-surfaces of vorticity have 
tubular shapes that are well-known from DNSs of fully developed turbulence. 
The heavy particles distribute themselves away from regions of high vorticity. 

We consider the motion of ten species of particles: tracers and nine heavy
particles, with different values of $\St$. We inject $\Np$ particles of each
species into the flow.  We collect data for averages after the system of
particles and the flow have reached a non-equilibrium, turbulent, but
statistically steady, state. 
It has been already observed by over laying positions of heavy particles 
on two-dimensional contours of vorticity that the heavy particles distribute 
themselves away from regions of high vorticity. 
Here we look at a time-series of $R$ and $\Delta$ obtained along the
trajectory of a particle; a typical example of such a time series 
for a tracer is shown in \Fig{fig:timqr}.
 The intersection of any one of these curves with the black, horizontal line
indicates the migration of a particle from one region of the $Q-R$ plane to
another.

\subsection{Persistence times via $Q$ and $R$}
\label{chap2:perqr}

\begin{figure*}
\includegraphics[width=0.32\linewidth]{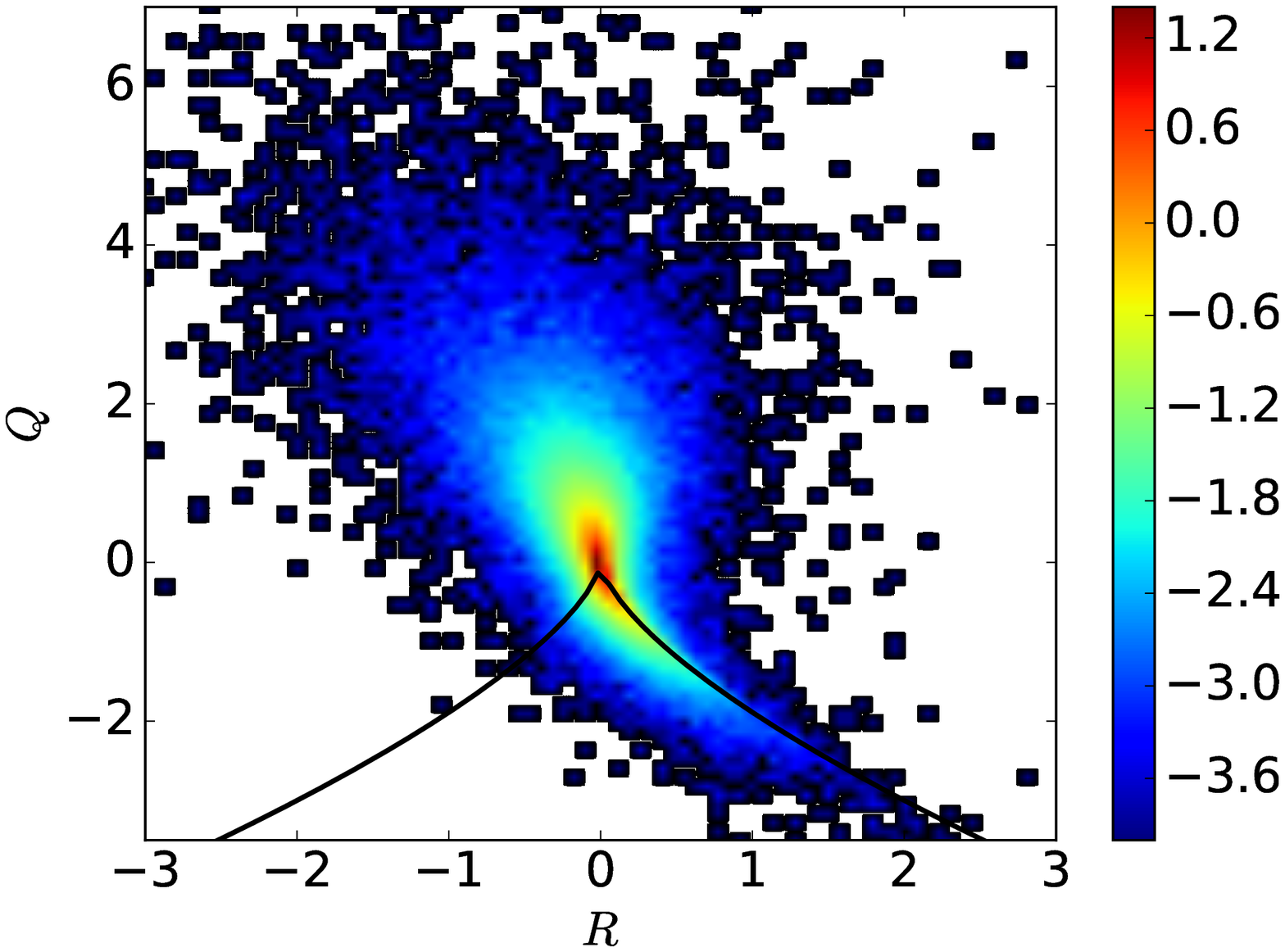}
\includegraphics[width=0.32\linewidth]{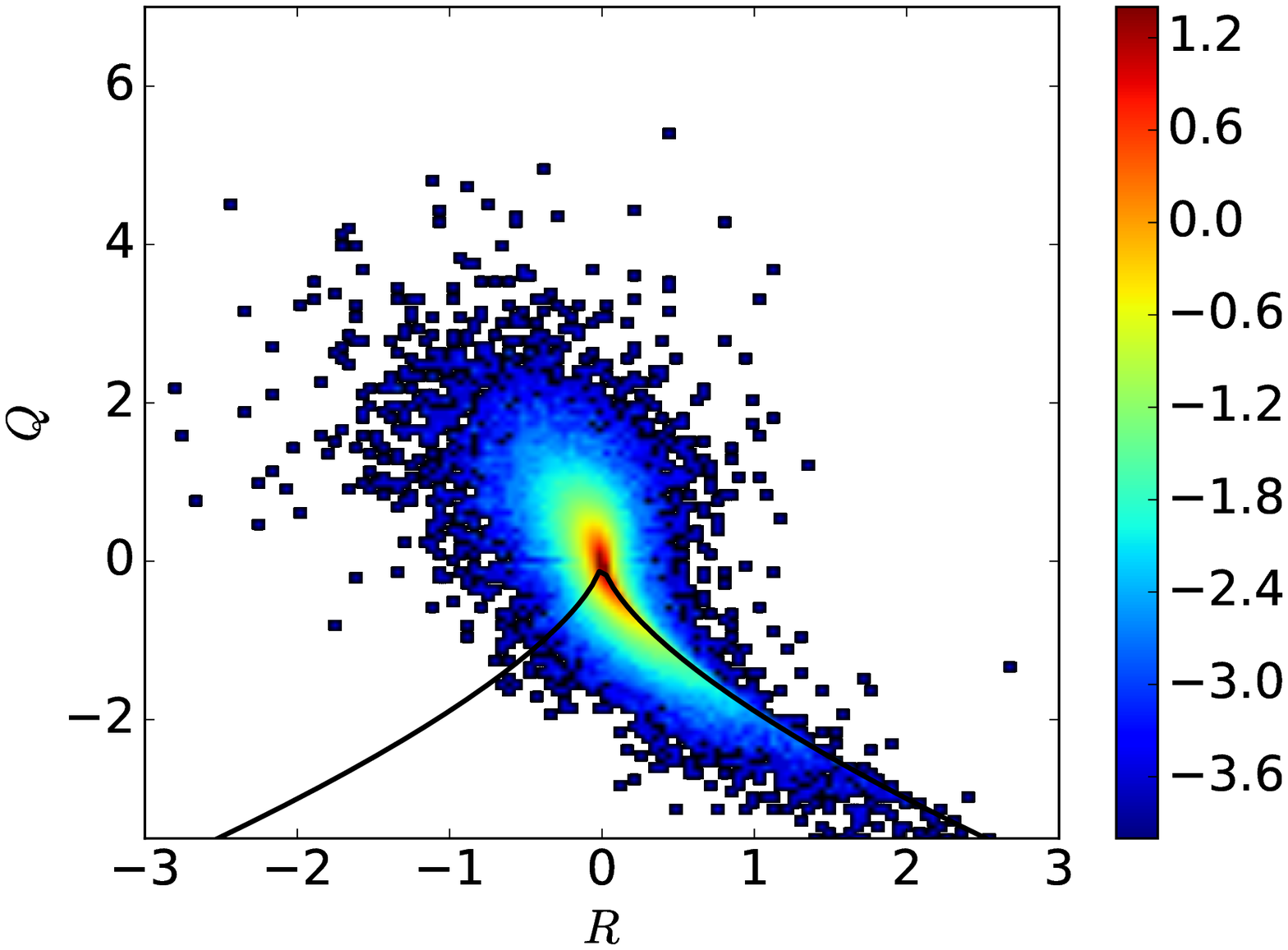} 
\includegraphics[width=0.32\linewidth]{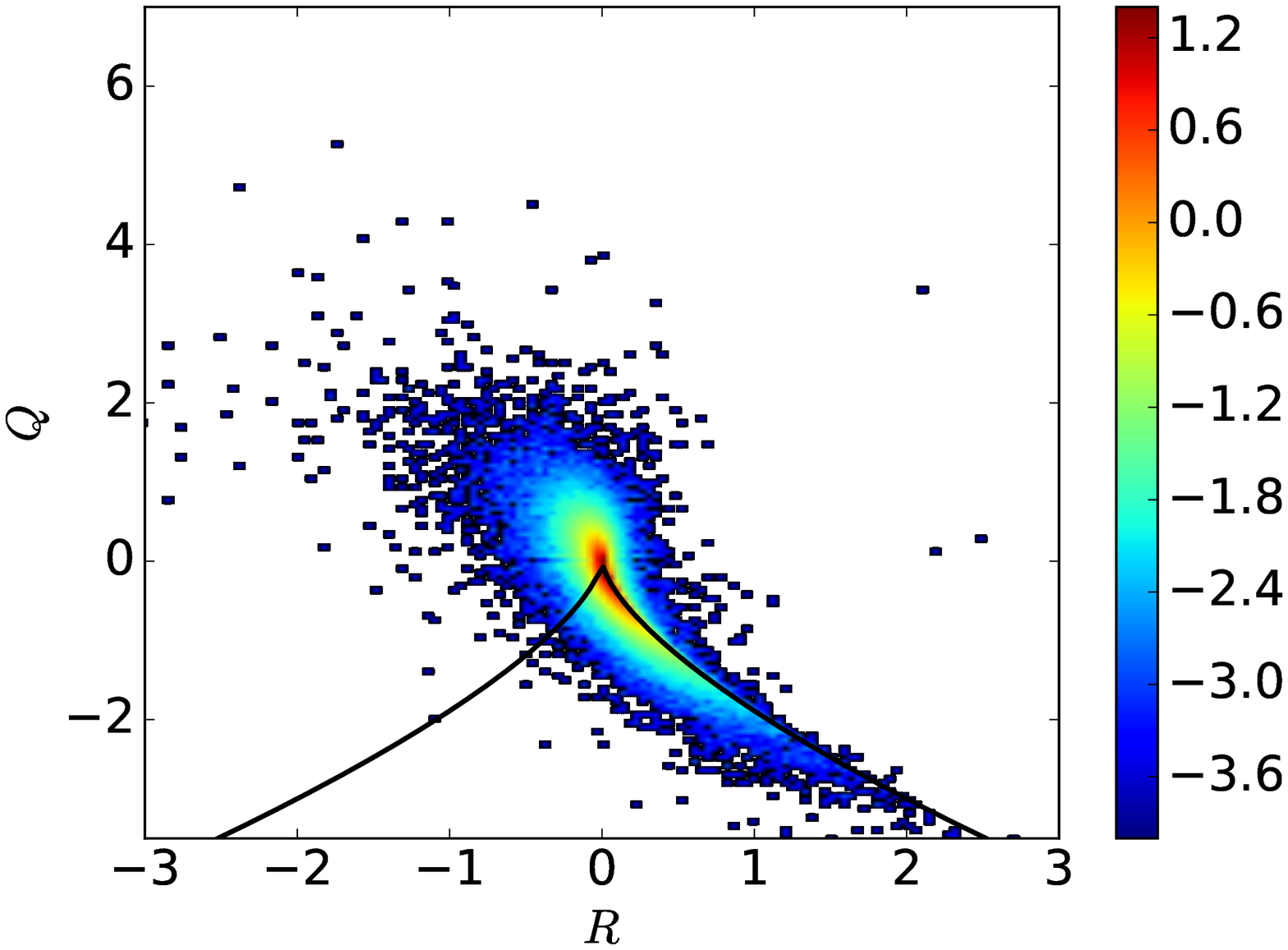}\\
\includegraphics[width=0.32\linewidth]{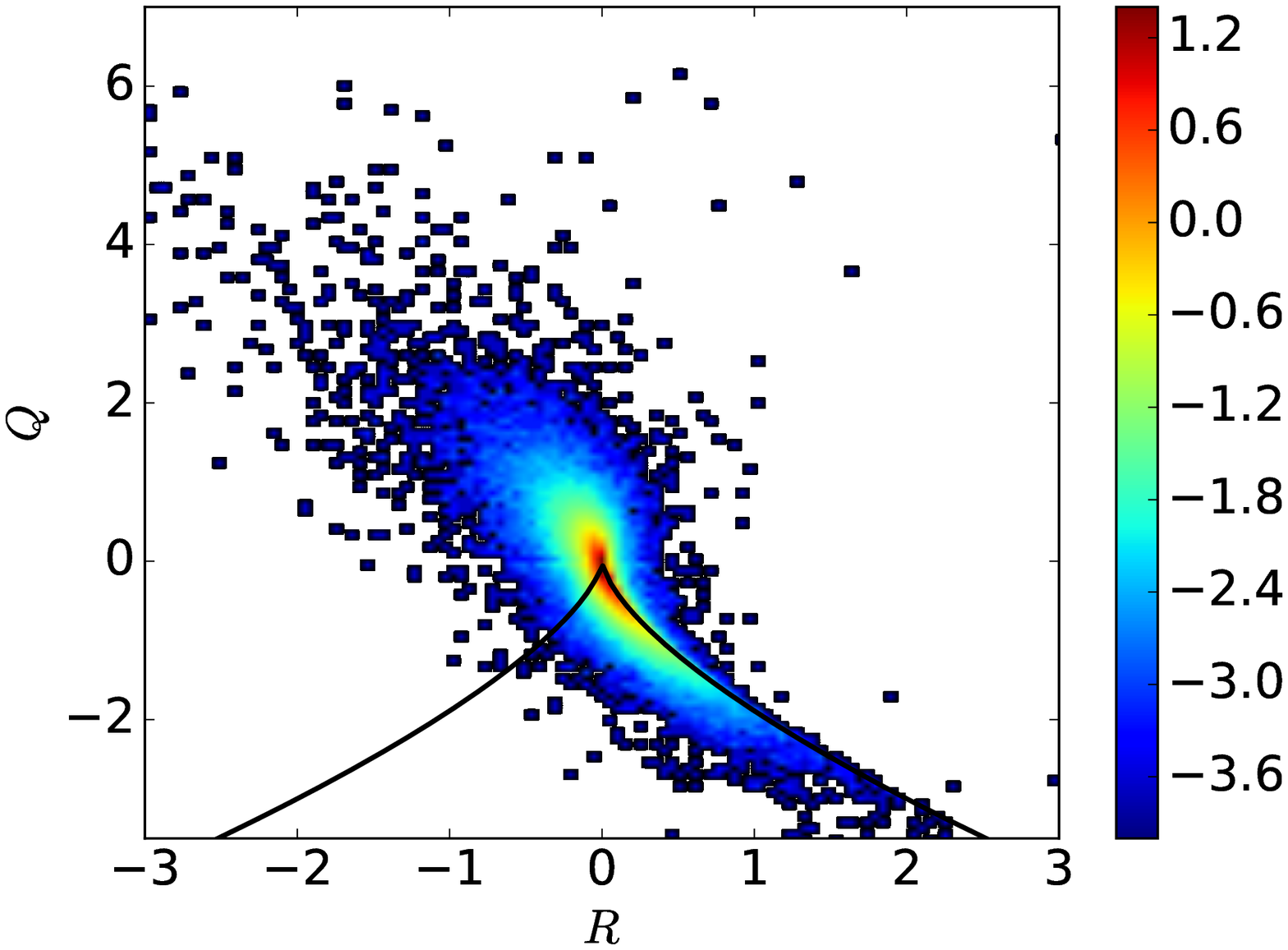}
\includegraphics[width=0.32\linewidth]{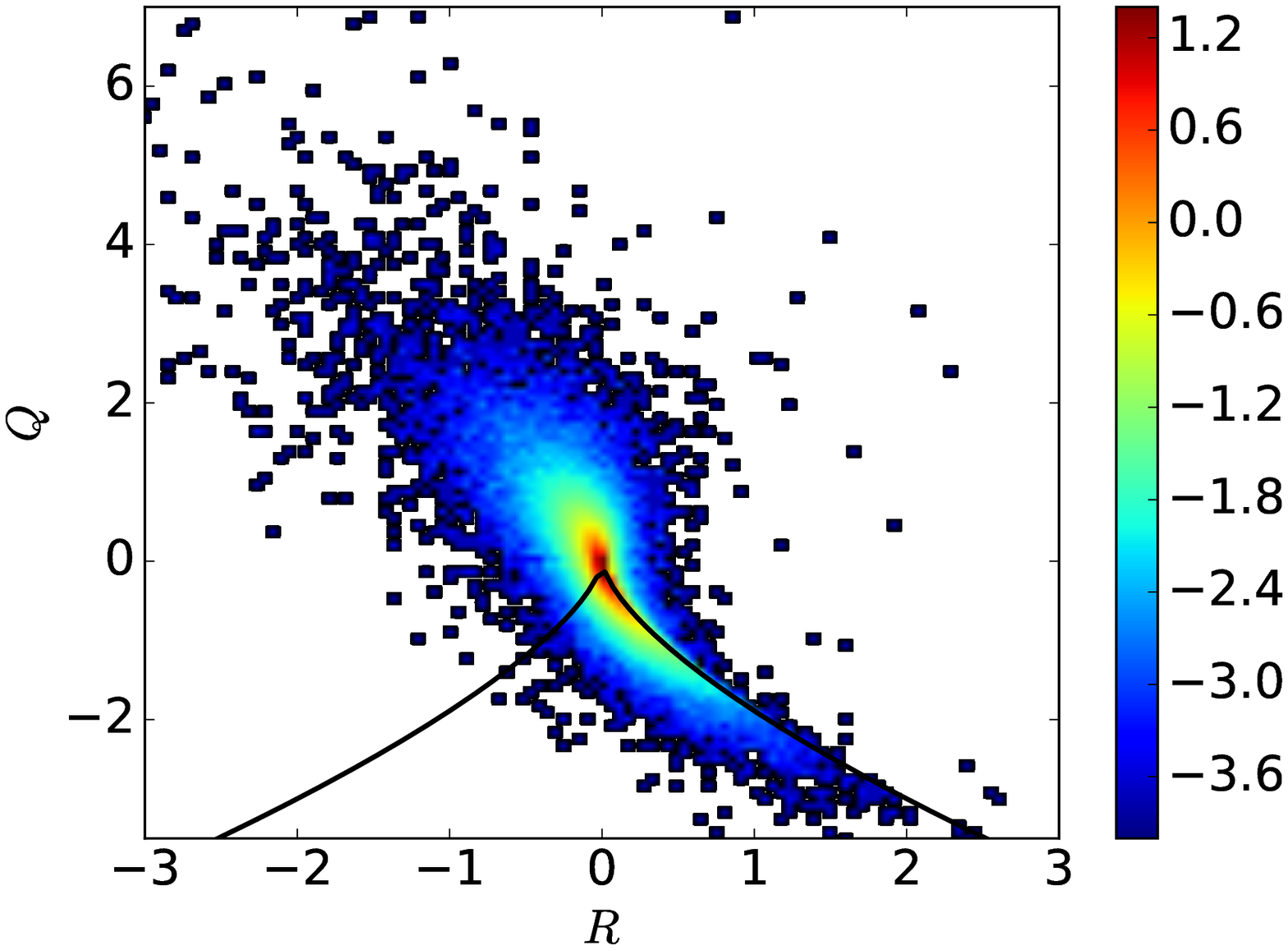}
\includegraphics[width=0.32\linewidth]{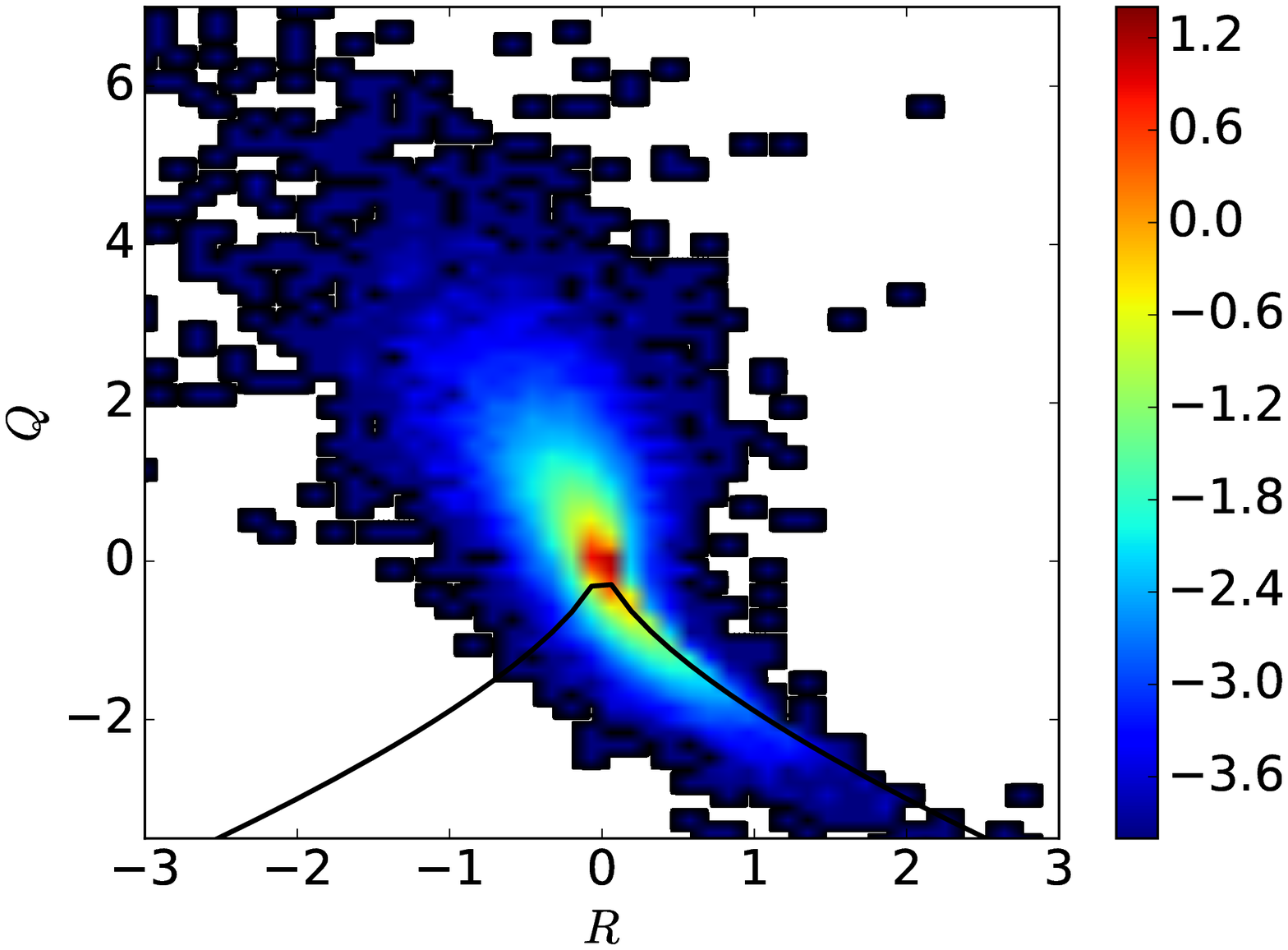}
\caption{(Color online) Contour plots of the joint PDFs of $Q$ and $R$, on log
scales, calculated along the trajectories of particles with different Stokes
numbers, from top the left corner, (a) tracers, (b) $\St=0.1$, (c) $\St=0.5$,
(d) $\St=1.0$, (e) $\St=1.4$, and (f) $\St=2.0$. $Q$ and $R$ are 
Nondimensionalized by $\Lambda^2$ and $\Lambda^3$, respectively, where
$\Lambda=u_\eta/\eta$. $\Delta=0$ curve is shown by solid black line, $\Delta>0$ 
corresponds to vorticity dominated region and $\Delta<0$ corresponds to strain dominated
region.} 
\label{fig:jpdf}
\end{figure*}

We follow the trajectory of each particle and calculate the components of $\AA$
and the values of $Q$ and $R$ at the particle position as a function of time.
In Fig.~\ref{fig:jpdf} we plot contours of the joint PDFs of $Q$ and $R$
$[P(Q,R)]$, on log scales; we calculate these values of $Q$ and $R$ along the
trajectories of tracers and heavy particles, for different values of $\St$.
These joint PDFs show that the tracers are more likely to be in
vorticity-dominated regions (region above the black curve in the $Q-R$ plane),
as compared to the heavy particles; in addition, the probability of finding
heavy particles in the vortical regions first decreases and then increases, as
we increase $\St$.     
\begin{figure*}
\includegraphics[width=0.49\textwidth]{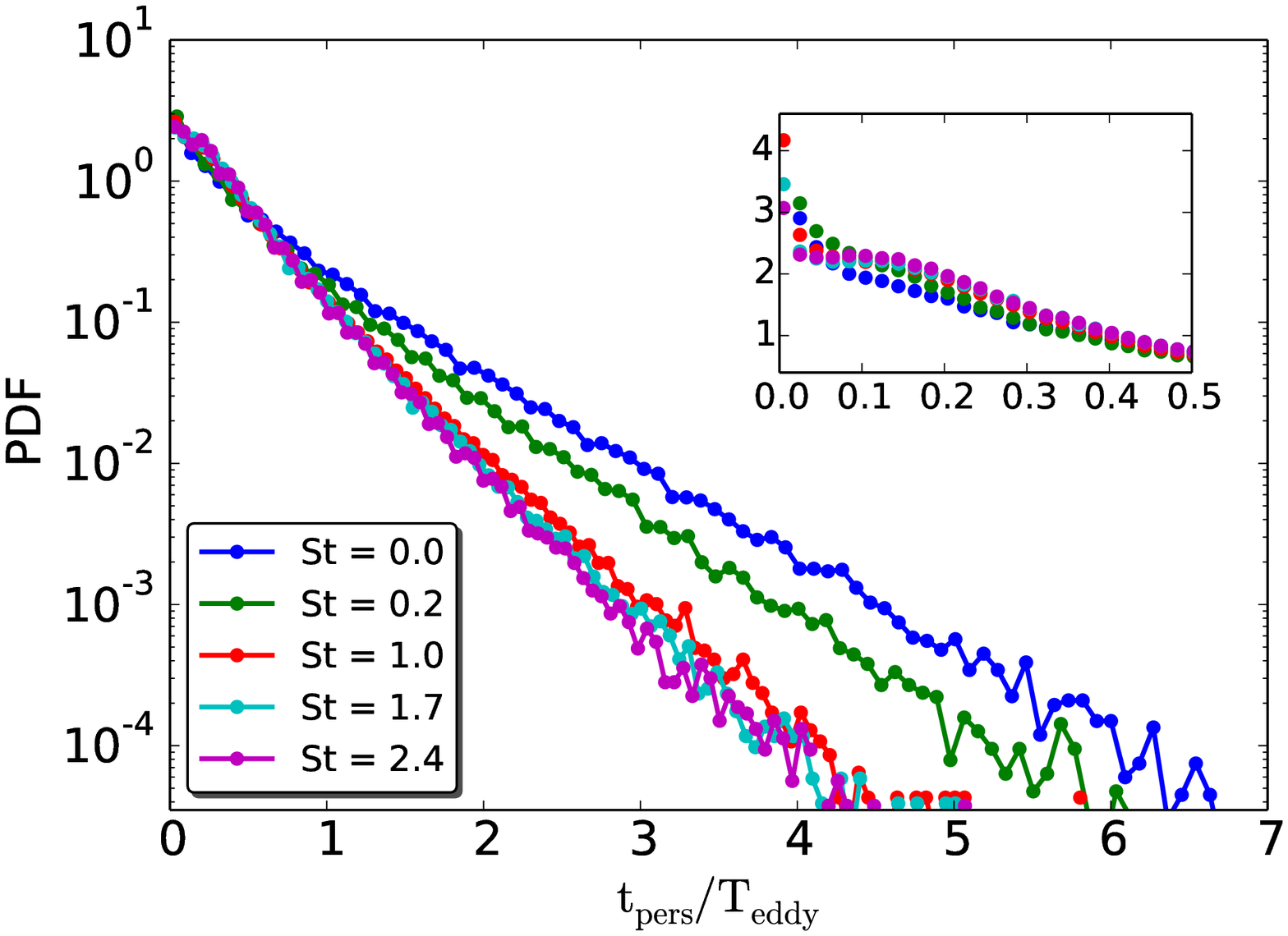}
\put(-170,150){\bf Region: B}
\includegraphics[width=0.49\textwidth]{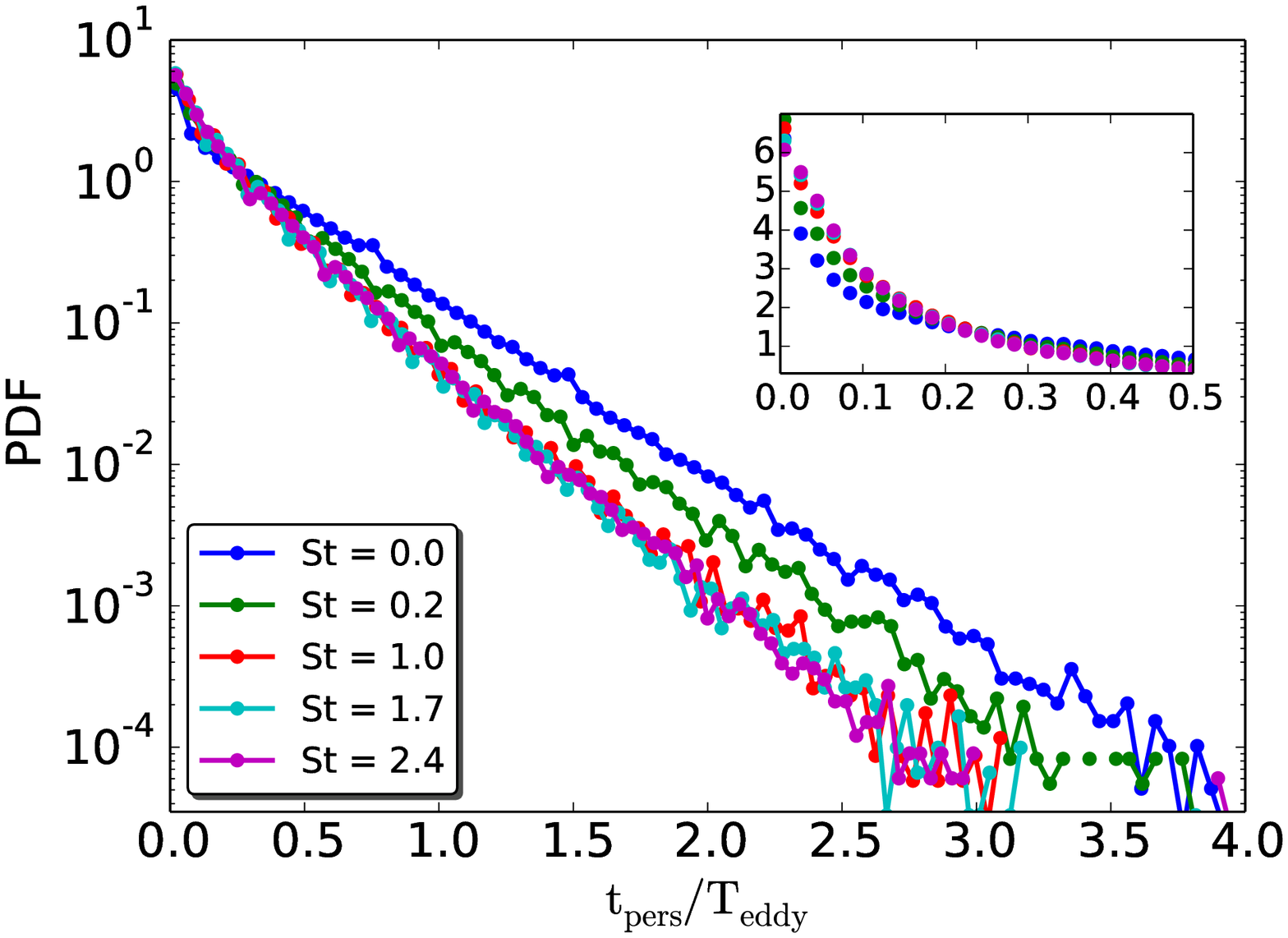} 
\put(-170,150){\bf Region: A}\\
\includegraphics[width=0.49\textwidth]{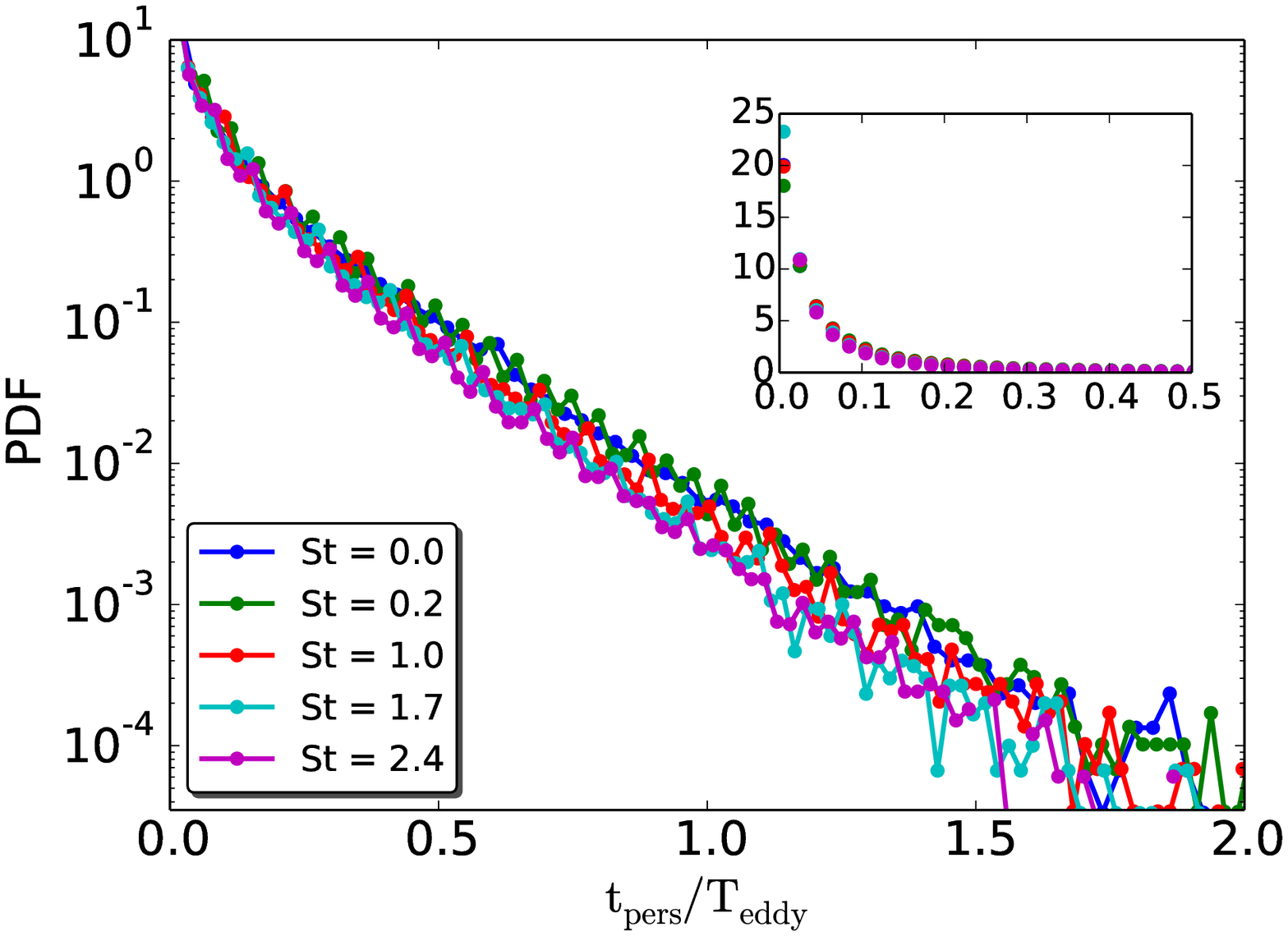}
\put(-170,150){\bf Region: C}
\includegraphics[width=0.49\textwidth]{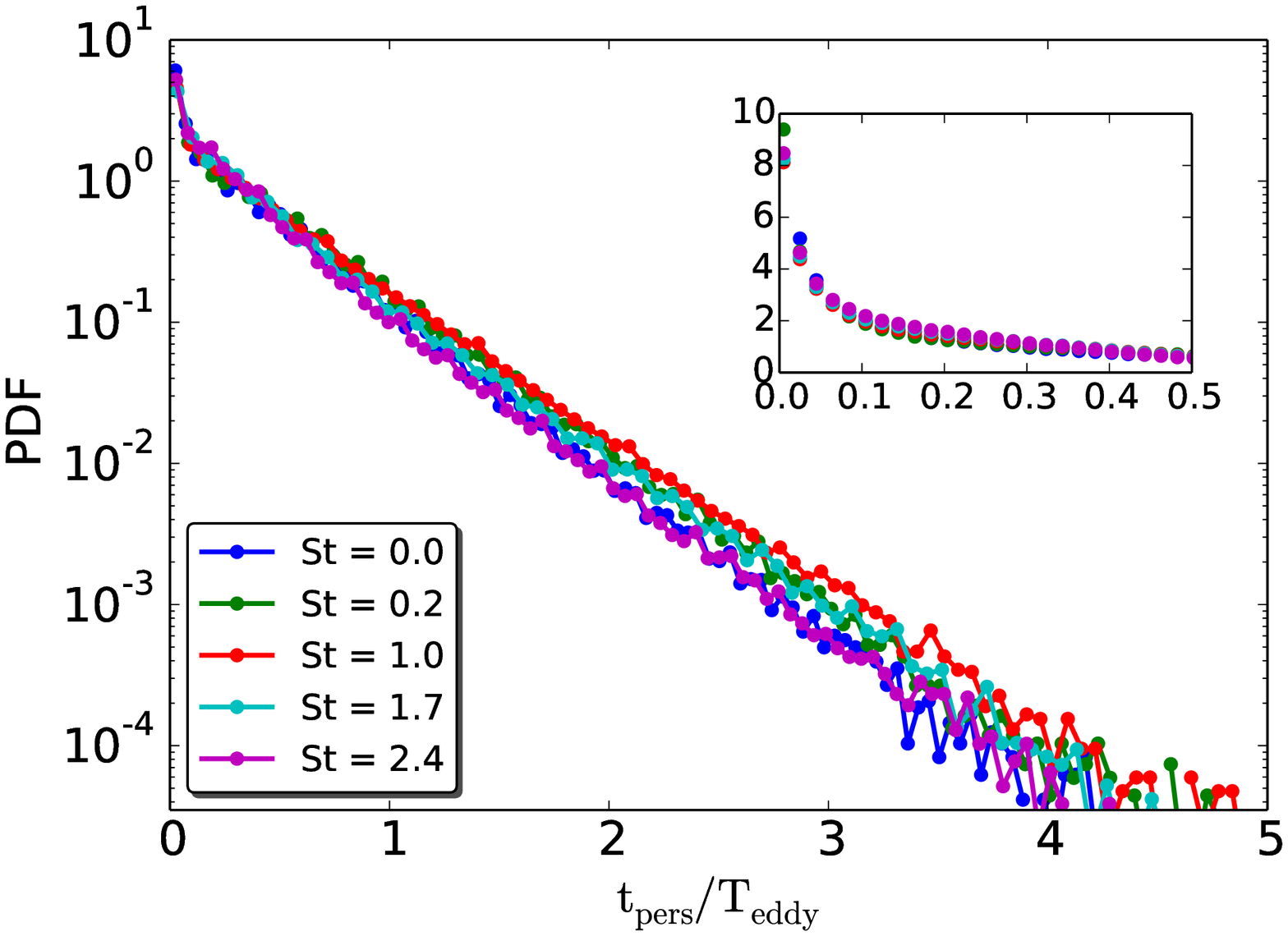}
\put(-170,150){\bf Region: D}
\caption{(Color online) Semi-log plots of the persistence-time PDFs
$\Pphi(\tper)$ of the times $\tper$ for the four regimes in the $Q-R$ plane,
for different Stokes numbers; the inset shows $\Pphi(\tper)$ for small
$\tper$.}
\label{fig:pdf_abcd}
\end{figure*}
\begin{figure*}
\includegraphics[width=0.49\textwidth]{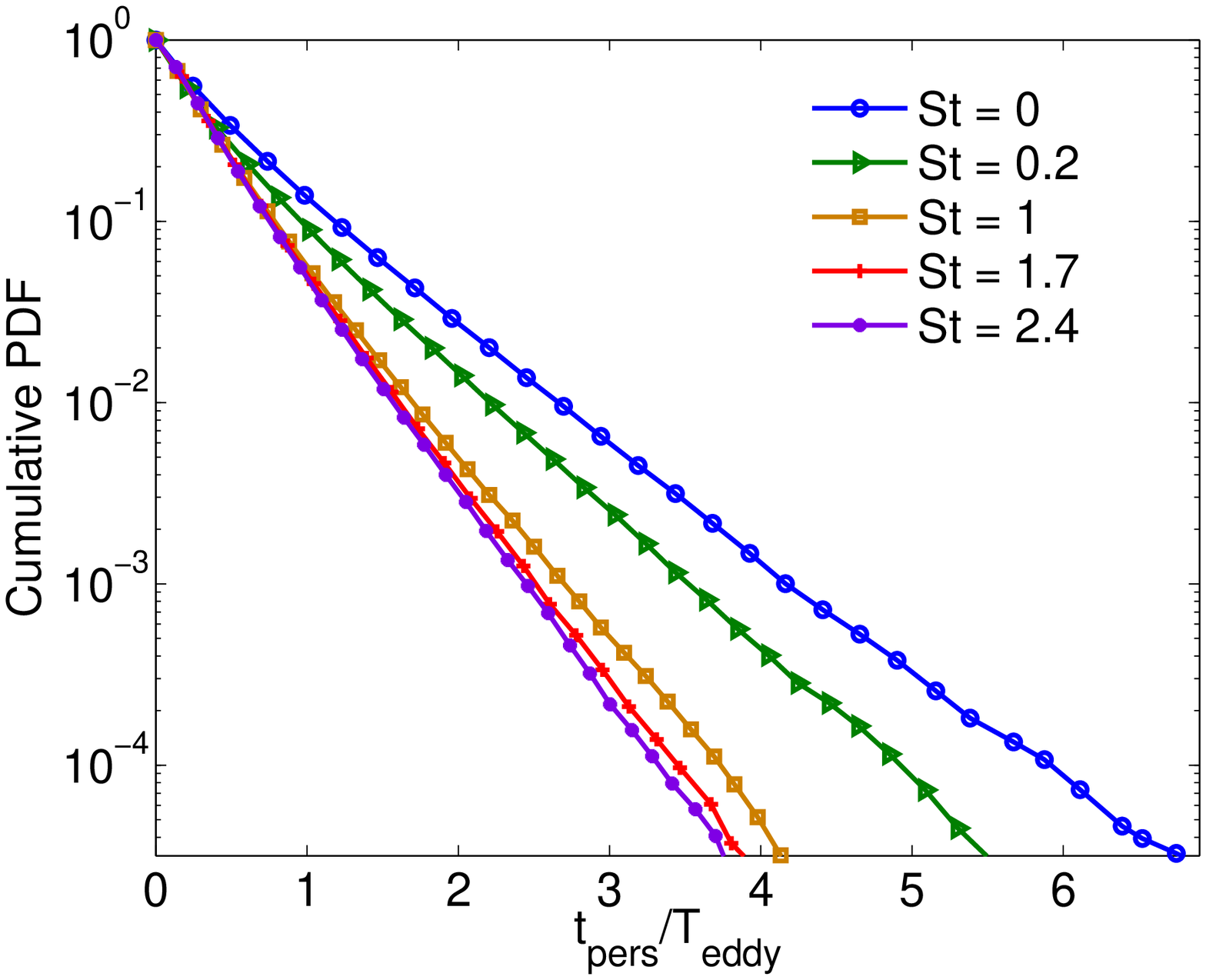}
\put(-170,40){\bf Region: B}
\includegraphics[width=0.49\textwidth]{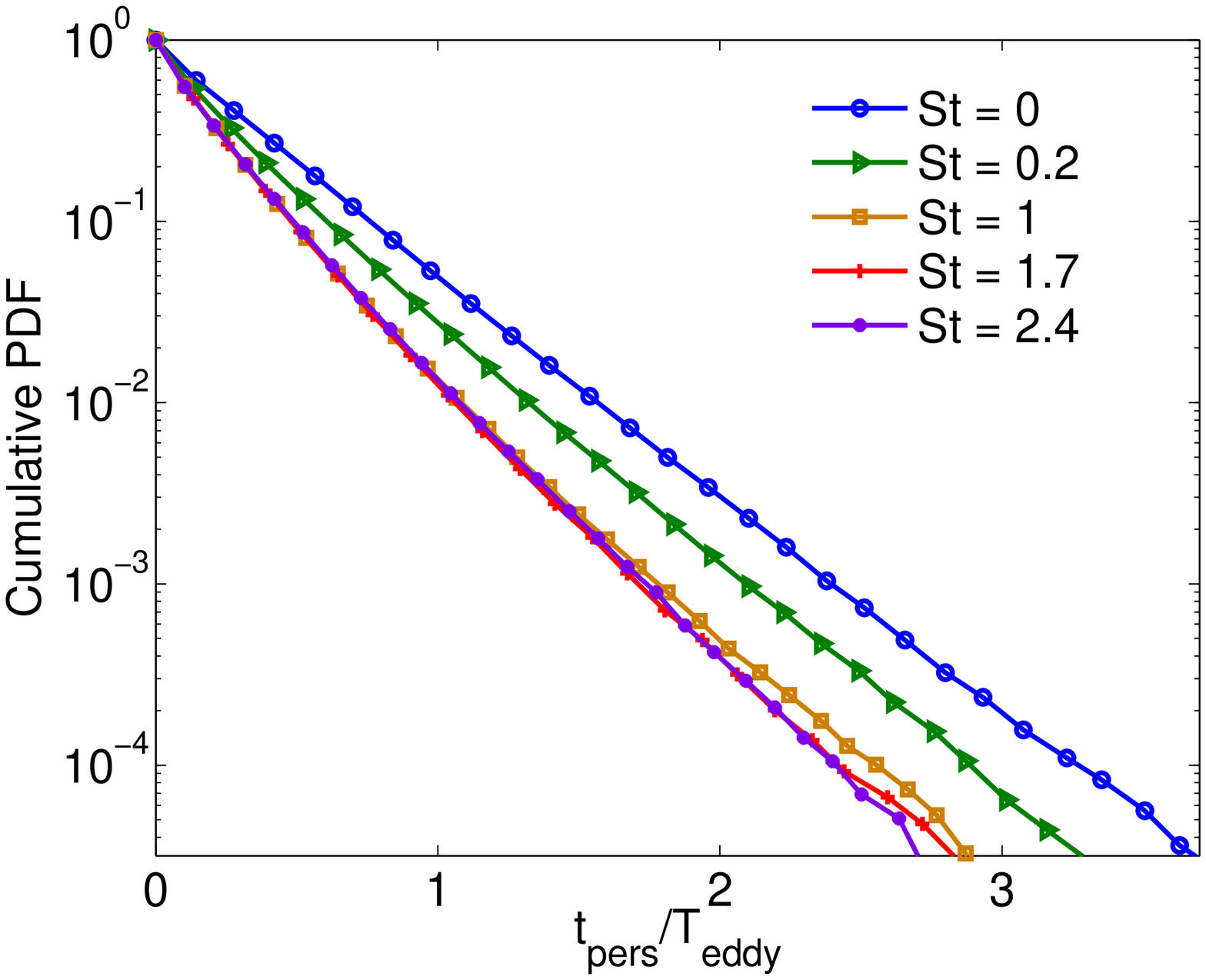}
\put(-170,40){\bf Region: A}\\
\includegraphics[width=0.49\textwidth]{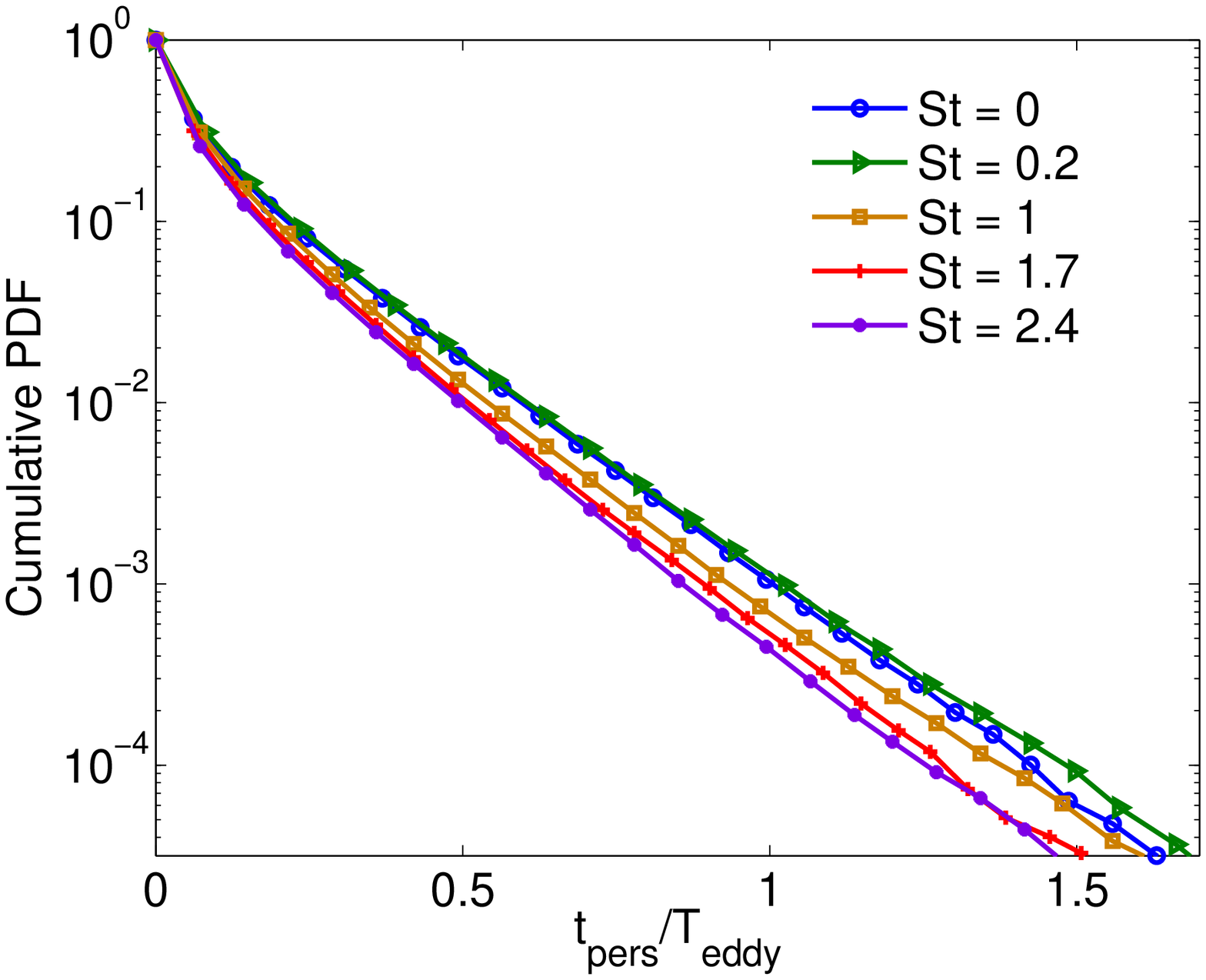}
\put(-170,40){\bf Region: C}
\includegraphics[width=0.49\textwidth]{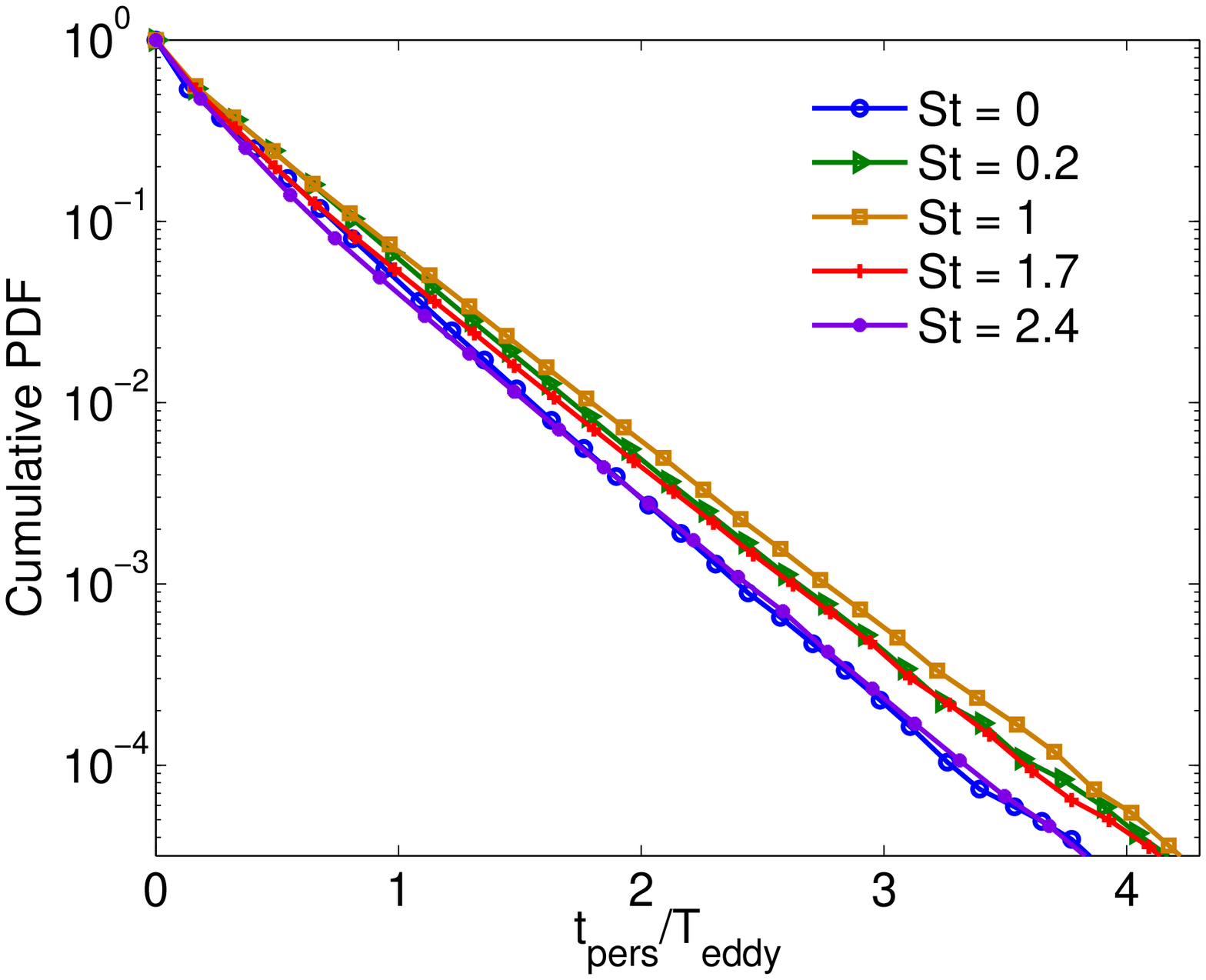}
\put(-170,40){\bf Region: D}
\caption{(Color online) Semi-log plots of the cumulative persistence time PDFs
(obtained by the rank-order method) for the four regimes in the $Q-R$ plane,
for different values of the Stokes number.}
\label{fig:qpdf}
\end{figure*}

We obtain the PDFs from our DNS as follows: (A) In the Eulerian framework, by
following the time evolution of $Q$ and $R$ at a fixed point $(x,y,z)$ in
space, we determine the time $\tper$ for which the flow at this point remains
in one of the four regions described above; (B) in the Lagrangian framework we
obtain the time  $\tper$ for which a tracer resides in one of these regions;
(C) the same calculation as in (B) but for heavy particles. For the Eulerian
PDFs we use a superscript ${\rm E}$, for tracer PDFs a superscript ${\rm L}$,
and for heavy-particle PDFs a superscript ${\rm I}$.  For each of the  four
regions in the $Q-R$ plane, we use the subscript A, B, C, and  D.  For example,
$\PI_{\rm A}$ denotes the PDF of times $\mathrm{t_{pers}}$ that a heavy
particle spends in the region A of the $Q-R$ plane.  

In Fig.~\ref{fig:pdf_abcd} we show semi-log plots of the PDFs of $\tper$ for
the four regions A, B, C, and D, which indicate that these PDFs display
exponentially decaying tails for large values of $\tper$. We give the forms of
these PDFs, for small values of $\tper$, in the insets (lin-lin plots). We find
that these PDFs do not go to zero as $\tper \to 0$. The qualitative natures of
these PDFs, for small $\tper$, are similar for regions A, C, and D, but not for
region B. These PDFs are obtained by computing the histograms and, therefore,
they suffer from binning errors. To overcome these errors, we calculate the
corresponding cumulative PDFs, by using the rank-order
method~\cite{mit+bec+pan+fri05}. We denote by $\QI_{\rm A}$ the cumulative PDF
(CPDF) that follows from $\PI_{\rm A}$; clearly, 
\begin{equation}
\PI_{\rm A}(\tper) \equiv \frac{d}{d\tper}\QI_{\rm A}(\tper) .
\end{equation}
In Fig.~\ref{fig:qpdf} we give semi-log plots of $\QI_{\rm A}(\tper)$, for
tracers and heavy particles, in regions A (top right), B (top left), C (bottom
left), and D (bottom right).  We observe that all these CPDFs have
exponentially decaying tails, from which we extract the characteristic time
scales $T_\alpha$ ($\alpha=$ A, B, C, or D) that we list in
Table~\ref{tab:talpha} for all species of particles. We also note that, in
regions A and B, which are vorticity dominated, $\TA$ and $\TB$ are largest for
tracers; and they decrease as $\St$ increases. Furthermore, for all species of
particles, $\TB > \TA$. The time scale $\TC$ for region C, which is
strain-dominated, does not change significantly with  $\St$.  The time scale
$\TD$ for region D, where axial strain dominates, assumes its lowest value for
tracers; and it changes only marginally as $\St$ increases. 

\begin{figure*}
\includegraphics[width=0.49\textwidth]{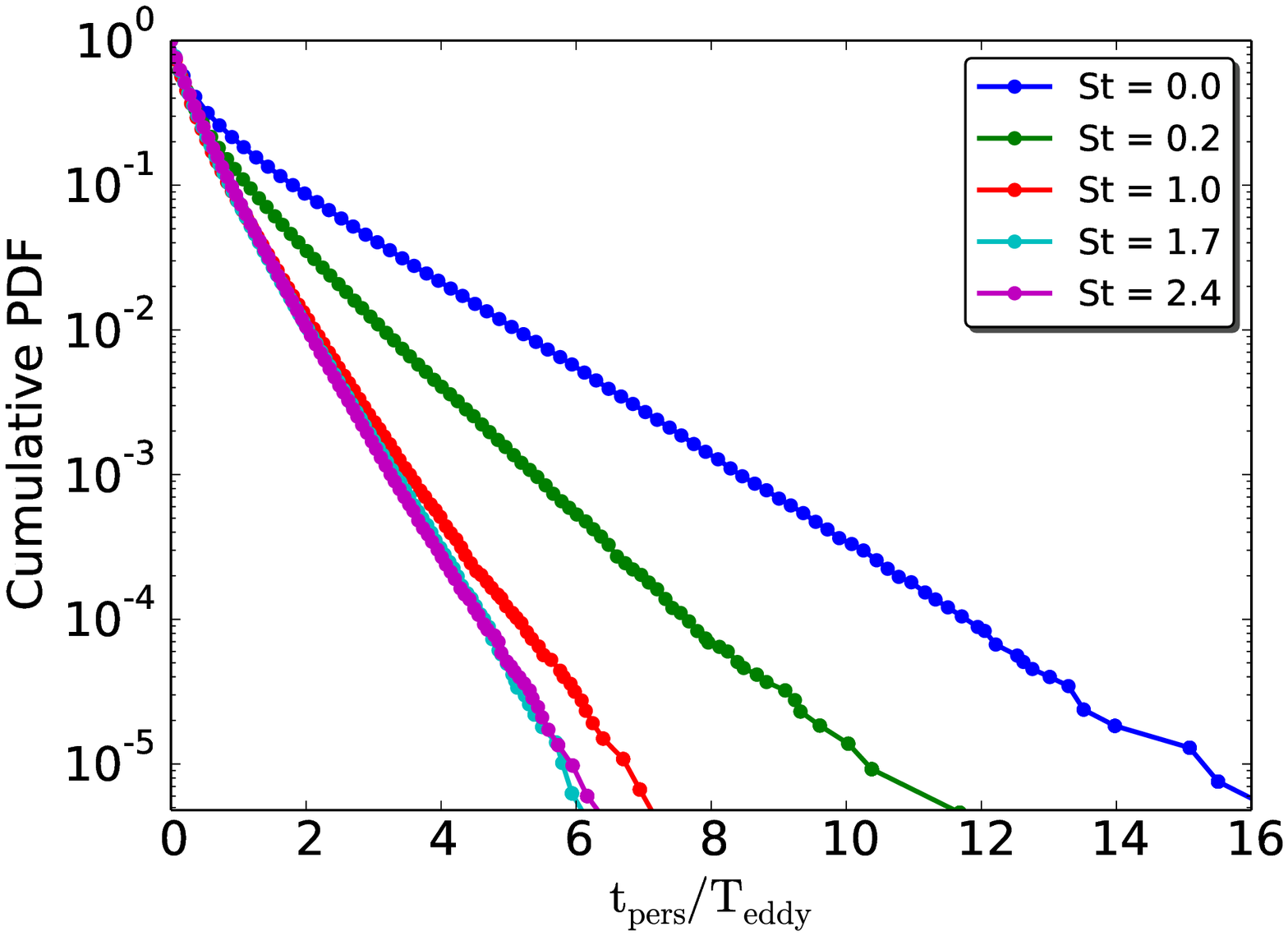}
\includegraphics[width=0.49\textwidth]{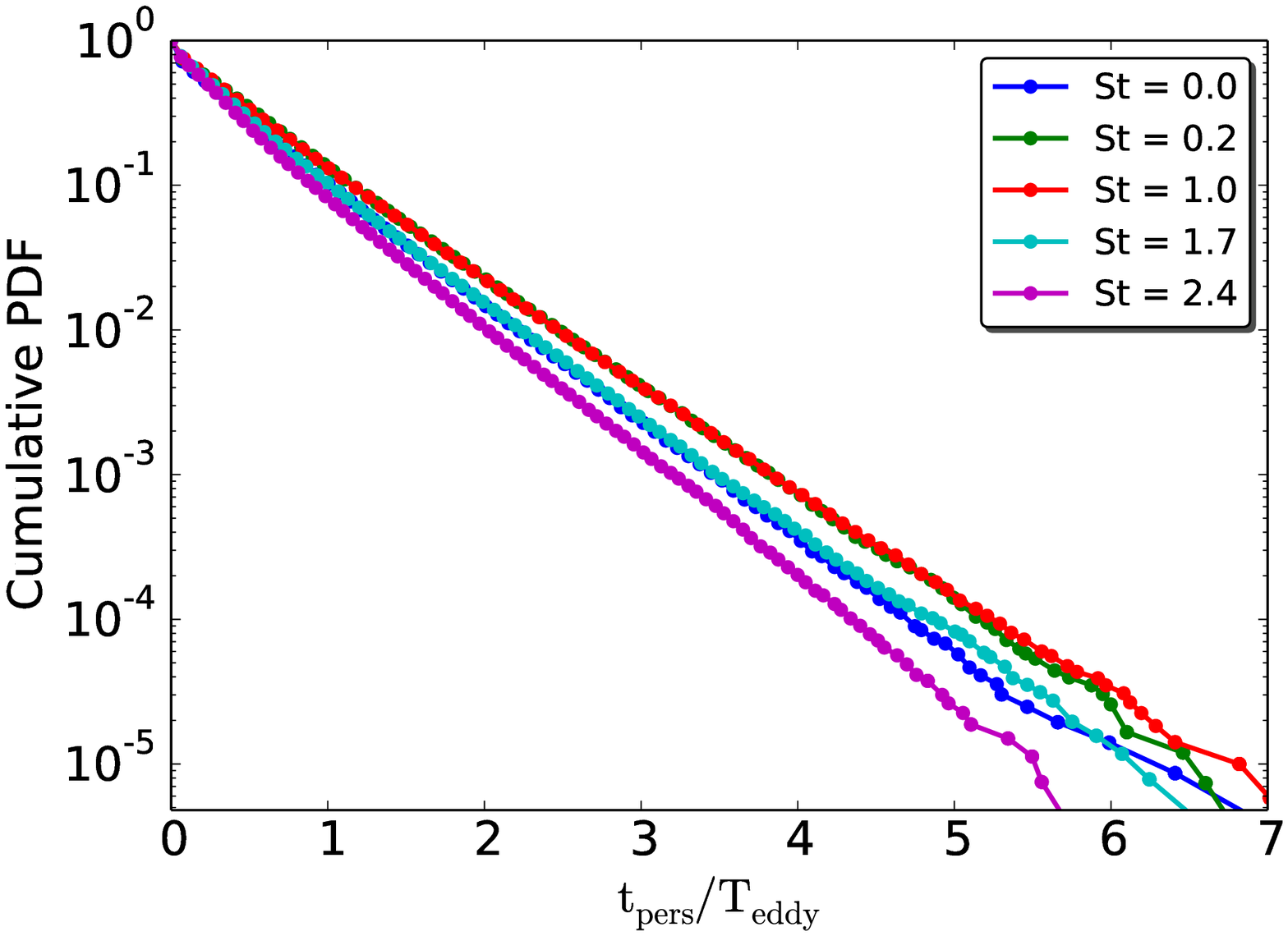}
\caption{(Color online) Semi-log plots of the cumulative persistence-time PDFs
(obtained by the rank-order method) for vortical ($\Delta > 0$, left panel) and
strain-dominated ($\Delta < 0$, right panel) regions, for different values of
the Stokes number $\St$ (the plot for tracers is labeled by $\St=0$). From the
slopes of the tails of these PDFs we extract the times $\TV$ and $\TS$, for
particle residence in vortical or strain-dominated regions of the flow,
respectively,.}
\label{fig:qpdf_delta}
\end{figure*}

To provide a clear answer to the question we pose in the title of this paper,
we must calculate the PDFs of the time $\tper$ for which heavy particles stay
in vortical regions of the flow. We do this by monitoring the sign of $\Delta$
along the trajectories of the particles, for $\Delta >0$ in vorticity-dominated
regions of the flow and $\Delta < 0$ in strain-dominated ones. In
Fig.~\ref{fig:qpdf_delta} we shows the CPDFs $\tper$ for the cases where
$\Delta$ remain positive (left panel) or negative (right panel), along the
trajectories of tracers ($\St=0$) or heavy particles; we find that these CPDFs
also have exponentially decaying tails.  We extract the time scales $\TV$ and
$\TS$, for particle residence in vortical or strain-dominated regions of the
flow, respectively, by fitting exponential functions to these tails. We list
these times in Table~\ref{tab:tvor_strain} for different values of $\St$. We
observe that $\TV$ decreases monotonically as $\St$ increases, whereas $\TS$
first increases and then decrease. Furthermore, the values of $\TV$ and $\TS$
indicate that tracers and heavy particles, with small values of $\St$, stay
longer in vortical regions of the flow than in strain-dominated ones, because
the difference between $\TV$ and $\TS$ is large here. By contrast, for heavy
particles, with high values of $\St$, the difference between $\TV$ and $\TS$ is
insignificant, so these particles spend roughly the same amount of time in
vortical regions of the flow as in strain-dominated ones.
\begin{table}
\begin{center}
\caption{Values of characteristic time scales, $T_\alpha$ for all four regions of $Q-R$
plane ($\alpha =$ A, B, C, D), calculated in the Eulerian frame for and in the frame of 
tracers and inertial particles, 
by fitting $Q_\alpha( \tper /T_{eddy})\sim \exp(-\tper/T_\alpha)$ 
to the cumulative PDFs of residence time. }
\begin{tabular}{l | c c c c}
\hline
           & $\TA/\Teddy$ & $\TB/\Teddy$ & $\TC/\Teddy$ & $\TD/\Teddy$ \\
\hline\hline
Eulerian    &		$0.13$  & $0.23$   &  $0.08$ & $0.13$ \\
Tracers			& 	$0.37$	& $0.68$	 &	$0.17$	 &  $0.37$ \\ 
$\St = 0.1$	&		$0.35$	& $0.63$	 &	$0.18$	 &	$0.39$ \\
$\St = 0.2$	&		$0.34$	& $0.58$	 &	$0.18$	 &	$0.41$ \\
$\St = 0.5$	&		$0.34$	& $0.48$	 &	$0.18$	 &	$0.42$ \\
$\St = 0.7$	&		$0.32$	& $0.45$	 &	$0.19$	 &	$0.41$ \\
$\St = 1.0$	&		$0.32$	& $0.44$	 &	$0.19$	 &	$0.42$ \\
$\St = 1.4$	&		$0.30$	& $0.41$	 &	$0.16$	 &	$0.42$ \\
$\St = 1.7$	&		$0.29$	& $0.39$	 &	$0.16$	 &	$0.41$ \\
$\St = 2.0$	&		$0.29$	& $0.39$	 &	$0.15$	 &	$0.42$ \\
$\St = 2.4$	&		$0.28$	& $0.37$	 &	$0.16$	 &	$0.39$ \\
\hline
\end{tabular}
\label{tab:talpha}
\end{center}
\end{table}
\begin{table}
\begin{center}
\caption{Values of the characteristic time scales, for the vortical ($\Delta >
0$) and strain dominated ($\Delta <0$) regions, calculated in the frame of
tracers and heavy particles for different values of $\St$.}
\begin{tabular}{l | c c }
\hline
           & $T_{\rm vortical}/\Teddy$ & $T_{\rm strain}/\Teddy$ \\
\hline\hline
Tracers			& 	$1.44$	& $0.54$	  \\ 
$\St = 0.1$	&		$1.11$	& $0.59$	  \\
$\St = 0.2$	&		$0.97$	& $0.59$	  \\
$\St = 0.5$	&		$0.73$	& $0.62$	  \\
$\St = 0.7$	&		$0.71$	& $0.63$	  \\
$\St = 1.0$	&		$0.64$	& $0.60$	  \\
$\St = 1.4$	&		$0.59$	& $0.59$	  \\
$\St = 1.7$	&		$0.56$	& $0.57$	  \\
$\St = 2.0$	&		$0.55$	& $0.55$	  \\
$\St = 2.4$	&		$0.55$	& $0.51$	  \\
\hline
\end{tabular}
\label{tab:tvor_strain}
\end{center}
\end{table}

\section{Conclusions}
\label{conc}

Our DNS of tracers and heavy particles in  statistically steady, homogeneous
and isotropic turbulence in the forced, 3D NS equation has helped us to explore
how long such particles spend in vortical regions of a turbulent flow and in
strain-dominated ones by combining properties of the velocity-gradient tensor,
which is well known in fluid mechanics, and the notion of persistence times,
which has received considerable attention in non-equilibrium statistical
mechanics. The $Q$ and $R$ invariants play a crucial role in our analysis of
PDFs and CPDFs of persistence times, conditioned on the values of $R$ and
$\Delta$. The exponential tails of these PDFs and CPDFs help us to extract time
scales that we identify with particle-residence times in vortical or
strain-dominated regions of the turbulent flow.  We hope that our detailed
study of persistence-time PDFs in 3D turbulent flows will lead to experimental
studies of such statistics for tracers and heavy particles. 

Our work is a natural generalization of a similar study for
tracers~\cite{perlekar2011persistence} in two-dimensional, statistically
steady, homogeneous and isotropic turbulent flows.  In two-dimensions, instead
of $Q$ and $R$, we must use the Okubo-Weiss parameter $\Lambda \equiv
\mathrm{det}(\AA)$.  This study has found that the PDF of the persistence time
$\tau$, for a Lagrangian particle in vortical regions,  displays a {\it
power-law tail}, i.e., $P^{\Lambda}(\tau_-) \sim \tau_-^{-\theta}$, where the
exponent $\theta \simeq 2.9$~\cite{perlekar2011persistence}.  By contrast, we
show that the residence-time PDFs in 3D turbulent flows display exponentially
decaying tails, for all species of particles and for all four regions in the
$Q-R$ plane. The most likely reason for this qualitative difference of
persistence-time PDFs (power-law as opposed to exponential tails) in 2D and 3D
fluid turbulence is that, in the 3D case, the velocity-gradient tensor $\AA$
always has one real eigenvalue, so tracers and particles can escape more
easily from vortical regions than they can in 2D turbulent flows. However, we
must also note that the extent of the power-law region seen in the 2D
study~\cite{perlekar2011persistence} increases with the Reynolds number.  The
Reynolds numbers that we can achieve in our 3D DNS is significantly lower than
that in 2D. Therefore, very-high-resolution, large-Reynolds-number DNSs of 3D
turbulence with tracers and particles are required to confirm the absence
of power-law tails in persistence-time PDFs here.

The clustering of heavy particles in 3D fluid turbulence has been characterized
by calculating a correlation dimension, which decreases first as $\St$
increases (for small $\St$), thus indicating clustering; but this
dimension reaches a minimum value near $\St \simeq 0.7$, and then increases to
a value $\simeq 3$ (i.e., a uniform distribution with insignificant clustering)
as $\St$ increases beyond $0.7$~\cite{bec2007heavy}. 
This can be understood in terms of singularities (caustics) in the 
(particle) velocity gradient field, see e.g., Refs.~\cite{pum+wil16,gus+meh16} for 
a review. The intuitive picture of clustering because of ejection from vortices
is not enough to understand the clustering.  
Nevertheless, we observe by plotting the joint PDFs of $Q$ and $R$ as measured along the
trajectories of heavy particles (Fig.~\ref{fig:jpdf}): the probability of
finding the heavy particles in the vortical regions first decreases and then
increases, as we increase $\St$. However, the characteristic times scales that
we have calculated for such particles in vortical structures behave differently,
insofar as they do not show such a clear, non-monotonic dependence on $\St$
(see Tables~\ref{tab:talpha} and \ref{tab:tvor_strain}).

Our DNS supports and quantifies the qualitative argument that heavy particles
spend less time than tracers in vortical regions in 3D turbulent flows.  However, the
residence time scales depend only weakly on $\St$, over the range we have in
Tables~\ref{tab:talpha} and \ref{tab:tvor_strain}. Surprisingly, these
characteristic times scales are comparable to the large-eddy turnover time. The
values of these time scales can be used as input parameters in developing a
model for the dynamics of the particles in turbulent flows.  If the same
characteristics time scales are calculated for Eulerian grid points, we find
that they are about one-tenth of the large-eddy turnover time, i.e., they are
of the same order as our Kolmogorov time scale. 
We see from the tails of the cumulative PDFs in Fig.~\ref{fig:qpdf_delta} 
that some of the particles can reside 
inside vortical regions for times that are much longer than $\Teddy$. 
Therefore, we need to run 
our DNSs for very long times to get good statistics. The results we present 
here have been 
obtained by running our DNSs for roughly $80\Teddy$. With such long runs, it is not possible 
to carry out very-high-resolution DNSs, at high Reynolds numbers and to 
obtain reliably 
the Reynolds-number dependence of persistence times.   

One of the many longstanding questions in turbulence concerns the lifetime of vortices. Clearly, to measure the 
lifetime of a vortex we must have a precise definition of a vortex, which is, in itself, still 
controversial, (see, e.g., Ref.~\cite{jeo+hus95}). One of the several different criteria used to define a vortex, 
called the Q-criterion, is precisely the condition $\Delta > 0$ that we have used. If we use this condition to 
define a vortex, then the time a tracer particle spends in a vortex can be considered as a measure of the 
lifetime of a vortex itself. Therefore, with this interpretation, we have provided an answer to the 
old question: What is the typical lifetime of vortical structures? The cumulative probability distribution 
of the lifetime of a vortex, in homogeneous and isotropic turbulence, given
in \Fig{fig:qpdf_delta}, has an exponential tail, which allows us to define a characteristic lifetime for a vortex; 
we give this lifetime in Table~\ref{tab:tvor_strain}. 
Other criteria for the definition of vortical regions can be used to measure the 
lifetime of vortices; and these may yield results that are different from those in Table~\ref{tab:tvor_strain}. 
An interesting 
attempt has been made to measure the PDF of the life time of vortical structures in Ref.~\cite{bif+sca+tos10} by using a 
DNS of light bubbles. This study lacked a precise definition of a vortex and had much smaller run times 
than those in our DNSs. Nevertheless, the characteristic lifetime of vortices, obtained in this study of 
Ref.~\cite{bif+sca+tos10}, are roughly equal to those we find.

An alternative way to define a vortical region (as opposed to a vortical point) 
is: ``to be a compact region of vorticity, possibly unbounded in one direction, surrounded by irrotational fluid. 
Strictly speaking, the viscosity has to vanish for this definition to make sense, but we suppose that 
the viscosity is very small, and we allow transcendentally small vorticity outside the
vortex ...'' 
(this quotation is from Ref.~\cite{pul+saf98}).  By using the lifetime of vortical regions, in a  model for vortex tubes, 
Mori~\cite{mor81} has argued that the characteristic dimension of vortical regions increases as a power-law in time, 
with a universal exponent equal to −3/2. As the Q-criteria is applicable to a point, but
not to a region, we cannot comment on this result.

\section{Acknowledgment}

This work has been supported  in part by Swedish Research Council under grant
2011-542 and 638-2013-9243 (DM), Knut and Alice Wallenberg Foundation (DM and
AB) under project ”Bottlenecks for particle growth in turbulent aerosols” (Dnr.
KAW 2014.0048), and Council of Scientific and Industrial Research (CSIR),
University Grants Commission (UGC), and Department of Science and Technogy
(DST India) (AB and RP). 
We thank SERC (IISc) for
providing computational resources. PP and RP thank NORDITA for hospitality
under their Particles in Turbulence program; DM thanks the Indian Institute of
Science for hospitality during the time some of these calculations were
initiated.  
\bibliographystyle{prsty}
\bibliography{ref2}
\end{document}